# Observing inside the coronagraphic regime with optimized single-mode nulling interferometry

E. Serabyn*[a], G. Ruane[a], D. Echeverri[b]

[a]Jet Propulsion Laboratory, California Institute of Technology, Pasadena, CA, USA 91109;
[b]Department of Physics, Mathematics and Astronomy, California Institute of Technology, Pasadena, CA USA 91125

**ABSTRACT**

The number of terrestrial exoplanets accessible to high-contrast coronagraphic imaging with large telescopes is limited by the smallest angular offset from bright stars at which coronagraphs can observe. However, it is possible to reach inside a telescope's coronagraphic regime by employing nulling interferometry across a telescope's pupil. Indeed, "cross-aperture" nulling interferometry can observe significantly closer to stars than typical coronagraphs, enabling observations even within the stellar diffraction core. Identifying an optimal nulling coronagraph, i.e., one with both a very small IWA and a high throughput for exoplanet light, would thus be of great interest. A systematic examination of available nulling options has therefore been carried out, which has led to three things. The first is a topological overview that unites both multi-aperture nulling interferometers and single-aperture phase coronagraphs into a common geometrical framework. The second is a new type of phase-mask coronagraph that has emerged from a gap in this framework, called here the "split-ring" coronagraph. The third is a clear identification of the optimal configuration for a nulling coronagraph, which turned out to be an aperture-plane phase knife, i.e., an achromatic π-radian phase shift applied to half the telescope pupil prior to focusing the telescope's point spread function (PSF) into a single-mode fiber. The theoretical peak efficiency of the phase-knife fiber coronagraph, 35.2% for a circular telescope aperture, is found to be almost twice that of the next most efficient case, the vortex fiber nuller, at 19.0%.

## 1. INTRODUCTION

While thousands of exoplanets have been discovered through the radial velocity and transit techniques, direct coronagraphic imaging has found far fewer, because of the demanding contrast requirements. However, a large enough space telescope that is both accurate and stable should be able to detect even terrestrial exoplanets around nearby solar-type stars[1,2]. Key to the success of such exoplanet space missions is the ability to suppress the very bright host starlight. Any coronagraph aiming to suppress starlight will of course have a minimum angular separation at which it can operate effectively, commonly referred to as the "inner working angle" (IWA). This is normally taken to be the angular separation from the star at which the throughput is half its large-angle asymptotic value, but some observations can also be carried out somewhat inward of this angle. As the population of small, rocky exoplanets is expected to rise toward smaller angular separations from host stars, the ability to reach the smallest possible angles is crucial to observing enough exoplanets to allow statistical analyses of their demographics, structural characteristics and compositions.

In this paper, we thus investigate the class of coronagraphs that possesses the smallest IWAs of all – nulling coronagraphs. In nulling interferometry[3], deep stellar cancellation is brought about by modifying the phase of one or more incoming light beams so as to center a dark (i.e., "null") interference fringe on a star, after which the resultant asymmetric point-spread function (PSF) can be spatially filtered to exclude the starlight. Indeed, it has been demonstrated that nulling coronagraphs can operate even within the core of the central stellar PSF[4]. As a result, here we have systematically modelled the throughputs of various single-mode (e.g., single-mode fiber coupled) nulling interferometers to find the most efficient variant. Upon consideration of the various nulling alternatives in the literature, not only was a clear winner found, that being the simple phase-knife nuller, but it was also found that the set of fixed-phase nullers correspond topologically with the set of phase-based single-aperture coronagraphs, which allowed the identification of an additional type of phase-mask coronagraph. All of these issues are explored in the following.

## 1.1 Background

The first single-telescope nulling observations[5,6] did not make use of spatial filtering, as the initial demonstrations and science observations did not require extremely high contrast. The two large dual-aperture nulling interferometers that followed used only very simple spatial filters, i.e., either a pinhole[7], or image plane filtering[8]. Nulling then returned to a single telescope aperture with the Palomar Fiber Nuller (PFN), which employed a single-mode fiber as both the nulling beam-combiner and a spatial filter, thus making it the first "photonic" nuller. By nulling a pair of subapertures across the Palomar Hale telescope, the single-baseline PFN was able[4] to null the primary star in a spectroscopic binary deeply enough to allow not only the direct detection of its secondary companion (with a brightness ~ 1% of the primary star) at a separation of ~ 1/3 of a diffraction beam width, but also the measurement of the diameter of the primary star itself. To illustrate the small IWAs reachable with such "cross-aperture" nullers, Fig. 1 compares the innermost (null) fringe of the Palomar cross-aperture nuller to the telescope's Airy pattern.

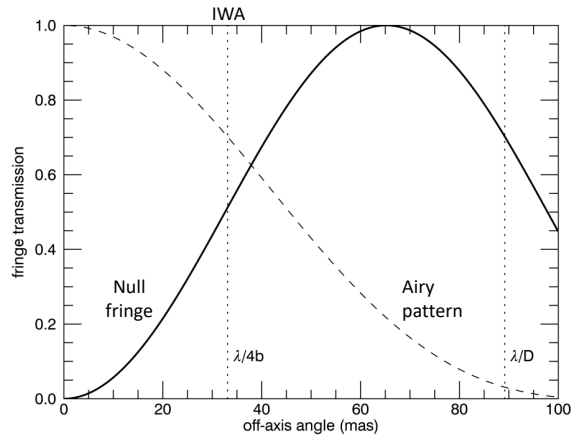

Figure 1. Comparison of the PFN's innermost null fringe to the Airy pattern of the full Hale telescope.

However, subaperture nullers with fixed phase shifts applied between subapertures are not the only possible nulling configuration, as the recently proposed vortex fiber nuller[9-11] (VFN) applies a continuous phase wrap around a single telescope pupil. While the VFN uses all of the telescope area, and so has a higher efficiency than a single-baseline subaperture nuller using small separated subapertures, it has a somewhat larger IWA. Likewise, the VFN and PFN have very different azimuthal responses. Such performance differences clearly suggest that a general examination of the tradeoffs between different nuller configurations would be very beneficial.

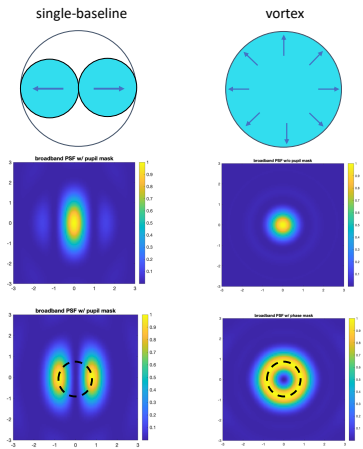

Figure 2. Comparison of fundamental nulling modes in Cartesian and cylindrical coordinates. Top row: pupil phases applied when nulling. Central row: PSFs with uniform phases applied. Bottom row: PSFs with nulling phases applied.

## 1.2 Pupil-plane nulling phase distributions in Cartesian and cylindrical coordinate systems

Deep cross-aperture nulling on a single-aperture telescope, as described here, consists of two essential steps – the application of a suitable phase pattern in the telescope beam's pupil plane, and the application of single mode filtering to the subsequent focal plane PSF. (Note that applying phase shifts to pupil subapertures prior to focusing the beam essentially gives a Fizeau beam-combiner configuration.) The two fundamental examples of such nulling are shown in Fig. 2, these being the simplest nulling cases in the Cartesian and cylindrical coordinate systems. In the former, a constant phase offset of $\pi$ radians is applied between two subapertures in the x-direction, while in the latter a continuous azimuthal phase wrap with a net $2\pi$ phase shift in a full circuit about the center is applied. The former yields linear fringes along the y direction, while the latter yields a ring-like fringe around the center. In both cases (Fig. 2), a spatially compact (point-like or line-like) central null is present, and the bulk of the bright starlight reaches the focal plane in the regions surrounding the central null. The starlight thus remains to be excluded, which calls for spatial filtering.

## 1.3 The role of a single-mode (SM) spatial filter

The central null in the two cases discussed above is mathematically zero only at a single point or line in the focal plane, implying that the null can only be deep over an extent the size of a single resolution element if the starlight immediately surrounding the central null is excluded from that resolution element. This issue is highlighted schematically in Fig. 2, in which the dashed lines indicate the rough extent of the central spatial mode (or resolution element). This region can be coupled to a spatial filter such as a SM fiber, if the fiber's single-mode acceptance angle is approximately matched to the telescope PSF core size. Because a SM fiber can internally propagate and externally couple to only a single symmetric spatial mode, the starlight surrounding the central null can be prevented from propagating in the fiber simply by making sure that the starlight arrives at the focal plane with an asymmetric field distribution, because an asymmetric field cannot couple to the symmetric fiber mode. Thus, deep nulling requires the arriving starlight field to satisfy

$$\vec{E}(-\vec{r}) = -\vec{E}(\vec{r}) \tag{1}$$

in the focal plane (at least within the acceptance cone of the SM fiber). This equation immediately implies that the incident focal-plane stellar electric field distribution must have an axial zero, i.e.,

$$\vec{E}(0) = 0, \tag{2}$$

and so (by virtue of the Fourier-transform relationship between focal-plane and pupil-plane fields) that the upstream pupil-plane field must have a zero-valued integral, i.e.,

$$\int_{pupil} \vec{E} \, dA = 0, \tag{3}$$

These conditions obviously apply to both of the basic nulling configurations illustrated in Fig. 2. Indeed, these two cases are conceptually equivalent to each other, being the simplest cases of nulling field distributions in their respective coordinate systems.

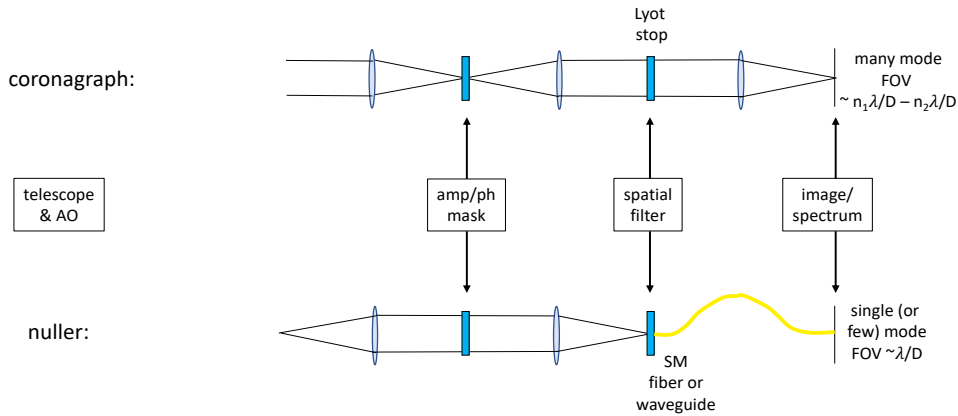

Figure 3. Comparison of coronagraph (top) and nuller (bottom) optical layouts, highlighting their functional similarities.

**1.4 The correspondence of nulling interferometer and phase coronagraph architectures**

Deep nulling thus involves two steps – the application of an appropriate phase pattern in the pupil plane, followed by spatial filtering in the subsequent focal plane. Note that this two-step process bears a striking similarity to phase-based coronagraphy, in which the phase pattern is instead applied in the focal plane, followed by a spatial filter (the "Lyot stop") in the subsequent pupil plane. To make this correspondence even more explicit, the optical layouts of the two systems are shown next to each other in Fig. 3, in which it can be seen that a phase pattern in one cardinal plane is followed by a spatial filter in its Fourier-transform plane.

This functional correspondence immediately suggests that cross-aperture nullers could be very simply implemented within existing coronagraphic benches by using the same optical layout, but inserting appropriate phase masks and spatial filters where needed. This allows future nulling interferometers to be envisioned not as separate instruments, but simply as additional modes of operation of coronagraphic benches. For completeness, we also note that this also implies that any of the previously-proposed Terrestrial Planet Finder-Interferometer[12] (TPF-I) and Darwin[13] nulling configurations (including even those that involve the added complexity of active phase shifting between pairs of different nullers) could be implemented relatively simply (compared to multi-aperture space-based interferometers) by making use of the requisite number of subapertures within a single telescope's downstream pupil-image plane.

## 2. MULTI-APERTURE NULLING CONFIGURATIONS

Many nulling configurations have been proposed over the years, especially in the era of the TPF-I and Darwin mission design studies[12,13], providing a large number of nulling configurations that could potentially be applied as subapertures of large single-aperture telescopes. These (Fig. 4) range from the simplest case of a single-baseline nuller, to linear arrays of larger numbers of telescopes[14,15] (that could be used either to generate higher-order stellar rejection or to actively phase modulate [i.e., "phase-chop"] between different independent nullers), to 2D arrays consisting of triangular[16,17], square[18], circular or elliptical configurations[13,19] with various numbers of apertures. Putting all or most of these configurations into some common framework before proceeding to evaluate a relevant subset would thus be desirable.

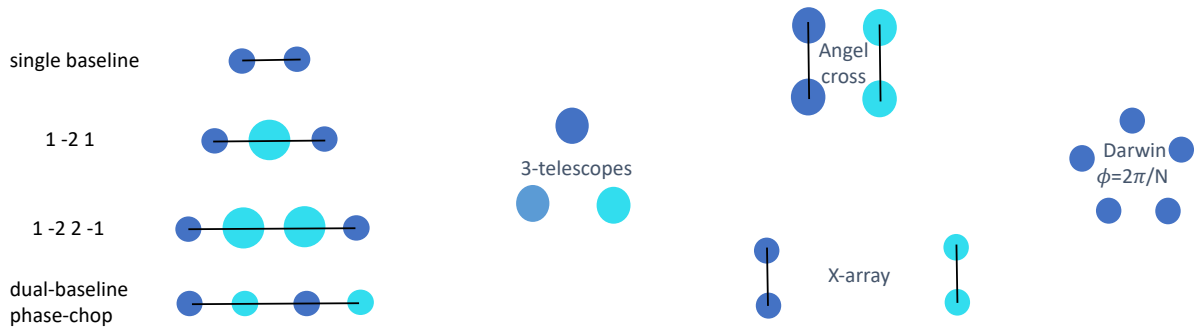

Figure 4. A (partial) catalog of nulling interferometry configurations from the literature.

To begin, we note that all of these nulling interferometer configurations are based on the application of appropriate phase shifts to different telescope beams. On the other hand, one can also note that the phase coronagraphs used behind a single-aperture telescope also rely on the application of an appropriate set of phase shifts, but in that case, applied to different sub-regions of a telescope's focal plane PSF. As both approaches rely on the use of relative phase, it makes sense to systematically compare them. Indeed, comparing the subset of fixed-phase nulling interferometers from Fig. 4 (i.e., those that are single nullers, and so do not involve phase-shifting between different nullers) to known (and in one case, unknown) phase coronagraph configurations immediately suggests a common topological framework for the two, which is illustrated graphically in Fig. 5. As can be seen in Fig. 5, this framework consists of taking each of the separate phase sections making up a given phase-coronagraph's focal-plane phase mask, separating them spatially, and distorting each subsection into the circular shapes assumed for separated-aperture nulling interferometry. Specifically, note that in each row of Fig. 5, the number of nulling apertures matches the number of coronagraphic phase-mask sub-sections, with matching phases and amplitudes in each (the latter assuming that the coronagraphic mask sub-region sizes are chosen appropriately). There is thus a one-to-one topological correspondence between the family of fixed-phase (i.e., non-modulated) separated-aperture (or separated subaperture) nulling interferometers, and the family of focal-plane phase coronagraphs.

However, note that to fill out this correspondence table completely, an additional type of phase coronagraph (seen on the right-hand side of the third row of Fig. 5) was needed to be invented, which we refer to hereafter as the "split ring" coronagraph. This coronagraph was needed to supply the correspondence to the higher order (1 -2 2 -1) nuller. In this coronagraph, each annular ring, including the central disk, is split in half, with phases of 0 and π radians in the two halves. The split-ring coronagraph can immediately be seen as an improved (i.e., less chromatic) version of Roddier's early phase disk coronagraph[20], in which a π phase shift was applied to an inner disk region of the PSF to cancel the outer part of the PSF. That system had the difficulty that the PSF's half-power radius is wavelength-dependent. However, by splitting both the central disk and the outer annulus in two here, and anti-phasing them, each ring (and the central disk) cancels itself, after which the outer ring can be used to further cancel the inner disk, thus reducing the importance of the chromatic nature of the radial phase transition. The split-ring coronagraph's performance will of course require further modelling, which is not appropriate here. Finally, note that although the line of the split can yield an undesirable linear feature in the focal-plane light distribution, such a feature will be entirely irrelevant if a one-sided dark hole is generated.

Returning now to the question of searching for optimal nullers, the unifying geometrical framework for nullers and phase coronagraphs described above, based on the geometric translation and warping of subapertures, makes it much easier to select a subset of subaperture nullers to focus on here, as the phase-mask coronagraph architectures seen in Fig. 5 immediately suggest thinking in terms of circular phase patterns. And indeed, most of the simplest separated-aperture nulling configurations consist of a number, N, of apertures (or subapertures) arranged in a circle. This includes even the N = 2 linear, N = 3 equilateral-triangle, and N = 4 square cases, as circles can be drawn through all of these sets of apertures. Therefore, in the following we concentrate on modelling the general case of the "N-subaperture circular nuller". We specifically consider N ranging from 2 to 6 here. However, note that our choice of circular nullers here is not entirely exclusive, as a pair of comparison cases are also included, those being the vortex nuller, in which a continuous phase wrap is applied around an aperture, making this essentially the limiting case of circular nullers for N → ∞, and the "phase-knife" nuller, in which a π phase shift is applied to half of the telescope aperture (making it the simplest version of the 2-subaperture nuller that could be implemented behind a single telescope, but which is not applicable to separate telescopes).

| Pupil-plane nulling interferometers | | Focal-plane phase coronagraphs |
|---|---|---|
| Single-baseline (-1 1) nuller | ○ ● | Phase knife |
| (-1 2 -1) nuller | ○ ● ○ | Roddier phase-disk (outer-ring conceptual split added here) |
| (1 -2 2 -1) nuller | ● ○ ● ○ | Combination of the above two: Split-ring coronagraph |
| Angel cross | ○ ● / ● ○ | 4-quadrant phase mask |
| N-aperture circular nuller (incl. multiple wraps) | ○ ● / ○ ● | Azimuthal staircase (includes 4QPM, 8OPM, digitized vortex) |
| (Continuous vortex for N → ∞) | ◌ ◌ | Continuous vortex mask |

Figure 5. Correspondence between separated-aperture nulling interferometer configurations and focal-plane phase-coronagraph configurations. Each black (white) region has a phase of 0 (π) radian; the grey regions have intermediate phases. The mottled areas bounded by dashed lines indicate negative phase regions. Note the outer region in Roddier's original phase-disk coronagraph was not divided into two halves; that division has been added here to illustrate the needed conceptual split of the outer region into two parts. The split-ring coronagraph in the third line is a generalized annular mask suggested here to complete the table. Note that both the ring-like and wedge-like cases can obviously be extended to higher numbers of segments.

Finally, before turning to our systematic comparison of N-subaperture circular nullers behind a single-aperture telescope, we briefly consider the simplest case of subaperture nulling using only 2 subapertures, to illustrate some of the issues and tradeoffs involved. As can be seen in Fig. 6, reducing the subaperture diameters from their maximum possible value (D/2) to their minimum (0) can bring an IWA decrease that is real but modest (maximally a factor of 2), but that this comes at the price of a rather extreme loss in collecting area, thus highlighting the fact that the factor that is actually important in coronagraphy is not the IWA itself, but the absolute transmission at the IWA, as that quantity could vary greatly, depending on a particular coronagraph's asymptotic transmission.

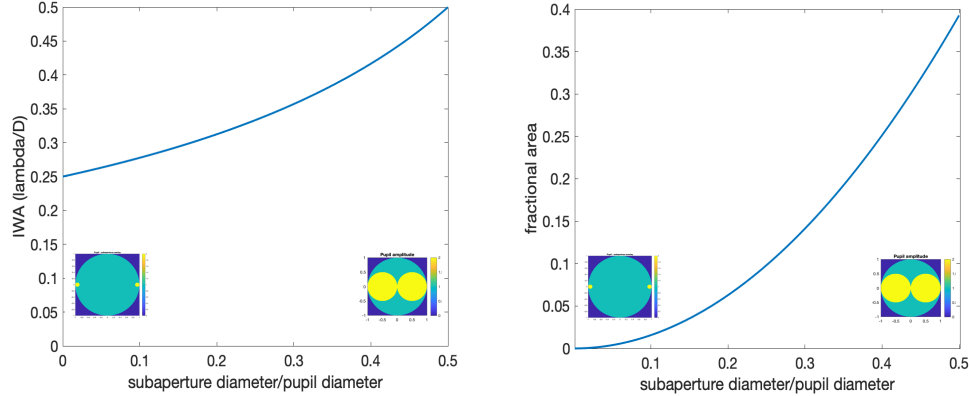

Figure 6. Left: IWA vs. subaperture diameter for maximally-spaced subapertures within a circular telescope pupil. Right: fractional telescope area used by a pair of maximally-spaced subapertures.

## 3. CIRCULAR NULLERS

### 3.1 The selected configurations and model outputs

In this section, we systematically explore the family of N-subaperture circular nullers. However, the number of free parameters available is still considerable, including the shape and size of both the telescope pupil and the subapertures within the pupil. To keep matters tractable, we aimed to use a single pupil shape, and in order to define this pupil, we first compared an ideal circular telescope pupil to the LUVOIR-B pupil[2] (which is segmented and has an irregular edge) for the case of a single-baseline subaperture nuller. Specifically, we calculated efficiencies (to be defined below) for a pair of subapertures having roughly 1/3 the diameter of the pupil, spaced as far apart as feasible, as in Fig. 7.

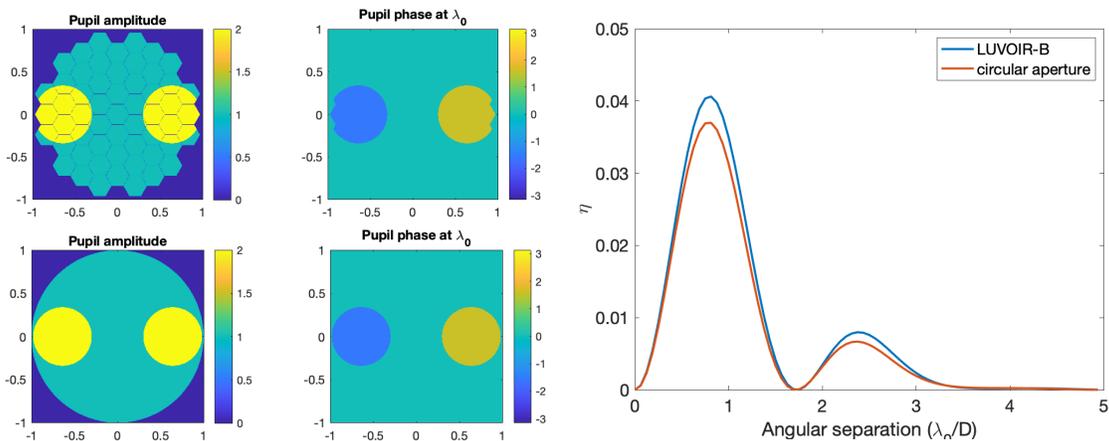

Figure 7. Top left pair: Amplitudes and phases of two circular mirror-symmetric nulling subapertures superposed on a LUVOIR-B pupil. Bottom left pair: The amplitudes and phases of two circular, maximally-spaced subapertures superposed on a circular telescope pupil. Right: Comparison of the efficiencies for these two cases.

As Fig. 7 shows, the two efficiency curves are very similar to each other, allowing us to focus on only one of these cases with no loss of generality. Hereafter, we thus exclusively use a circular telescope pupil. There is then no need to specify the actual physical pupil diameter, D, as all of our results are given in terms of $\lambda/D$, where $\lambda$ is the observation wavelength. However, another reason for selecting a simple circular telescope aperture here is that it allows a unique and well-defined means of selecting the subaperture shapes and sizes. In particular, for each choice of N, we select equal round subapertures with radii that are the maximum subaperture radius, r, that would allow fitting the required number, N, of such subapertures into an unobscured circular telescope aperture of radius R. Simple geometry then gives

$$r = \frac{R}{1+\frac{1}{\sin\left(\frac{\pi}{N}\right)}}, \qquad (4)$$

and a distance for the centers of the subapertures from the center of the telescope pupil of

$$R - r = \frac{R}{1+\sin\left(\frac{\pi}{N}\right)}. \qquad (5)$$

For this comparison, we calculated (in Matlab) several quantities for each subaperture configuration (i.e., for the circular N = 2 to 6 configurations, and for the other two configurations mentioned), those being the focal-plane PSF, and the resultant radial and azimuthal efficiency curves, where efficiency refers to the fraction of the exoplanet light incident on the telescope that makes it through the single mode fiber to the detector. With larger numbers of subapertures, greater numbers of unique phase wraps are possible (i.e., reaching $2\pi n$ in a complete circuit about the center, where $1 \leq n \leq N/2$), and so these outputs were calculated for all possible unique phase wraps. Finally, we approximately matched the fiber-mode width to the telescope PSF's FWHM, and did not vary that quantity here. The model results for N = 2 through 6 are presented in Figs. 8-12, for the vortex nuller in Fig. 13, and for the phase-knife nuller in Fig. 14.

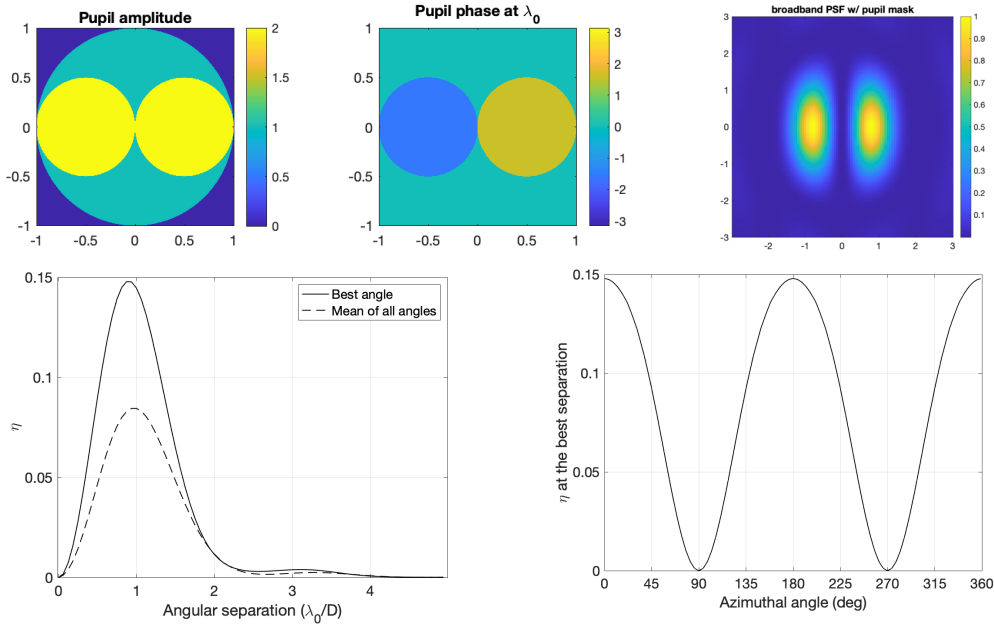

Figure 8. 2-subaperture nuller. Top row: pupil amplitudes and phases, as well as the resultant focal-plane PSF for a pair of maximal subapertures. Bottom row: the radial and azimuthal efficiencies for these subapertures.

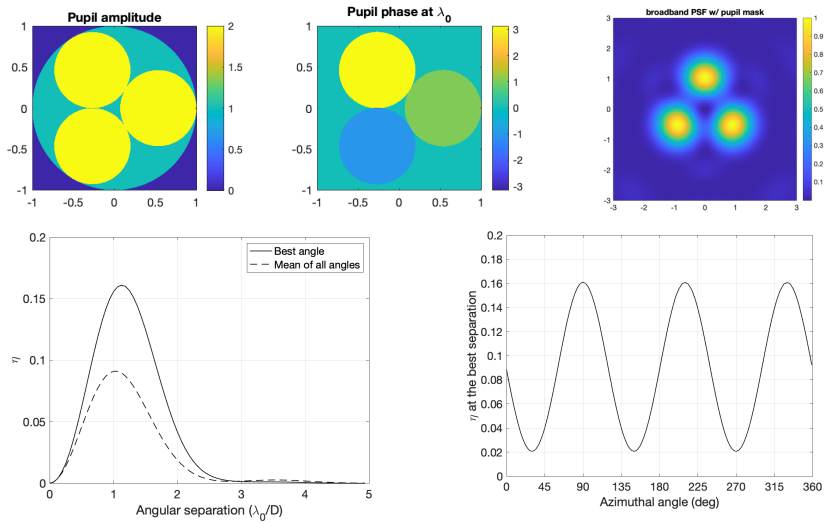

Figure 9. 3-subaperture nuller. Top row: pupil amplitudes and phases, as well as the resultant focal-plane PSF for 3 maximally-spaced subapertures. Bottom row: the resultant radial and azimuthal efficiencies.

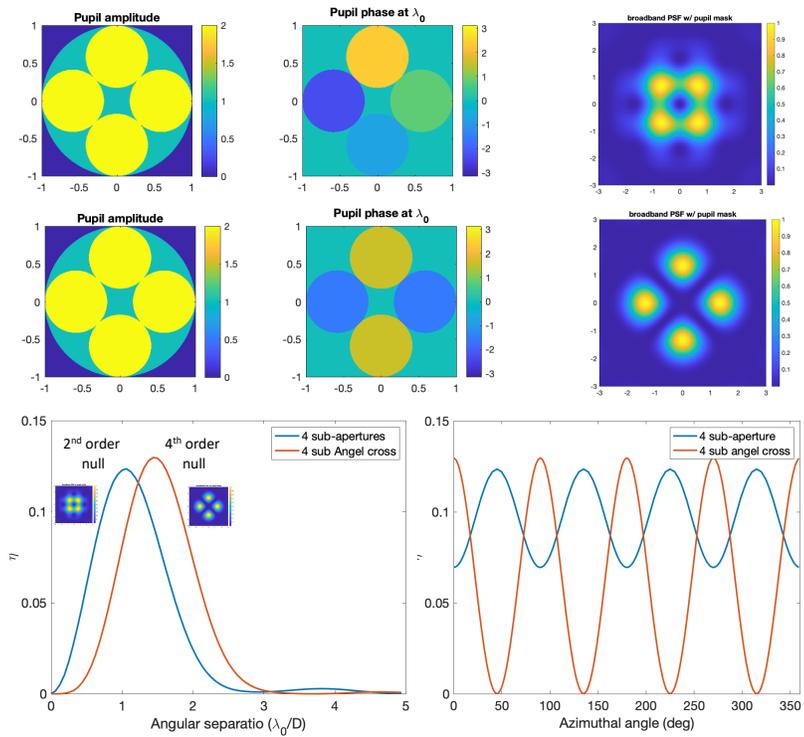

Figure 10. 4-subaperture nullers. Top row: pupil amplitudes and phases, as well as the resultant focal-plane PSF for a total phase wrap of $2\pi$ radians for 4 for maximally-spaced subapertures. Middle row: pupil amplitudes and phases, as well as the resultant focal-plane PSF for a total phase wrap of $4\pi$ radians. Bottom row: the resultant radial and azimuthal efficiencies.

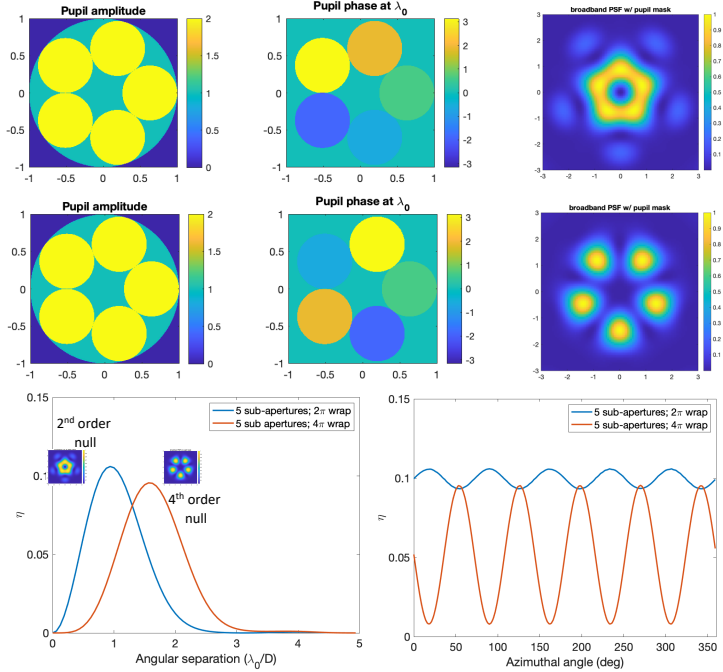

Figure 11. 5-subaperture nullers. Top row: pupil amplitudes and phases, as well as the resultant focal-plane PSF for a total phase wrap of $2\pi$ radians for 5 for maximally-spaced subapertures. Middle row: pupil amplitudes and phases, as well as the resultant focal-plane PSF for a total phase wrap of $4\pi$ radians. Bottom row: the resultant radial and azimuthal efficiencies.

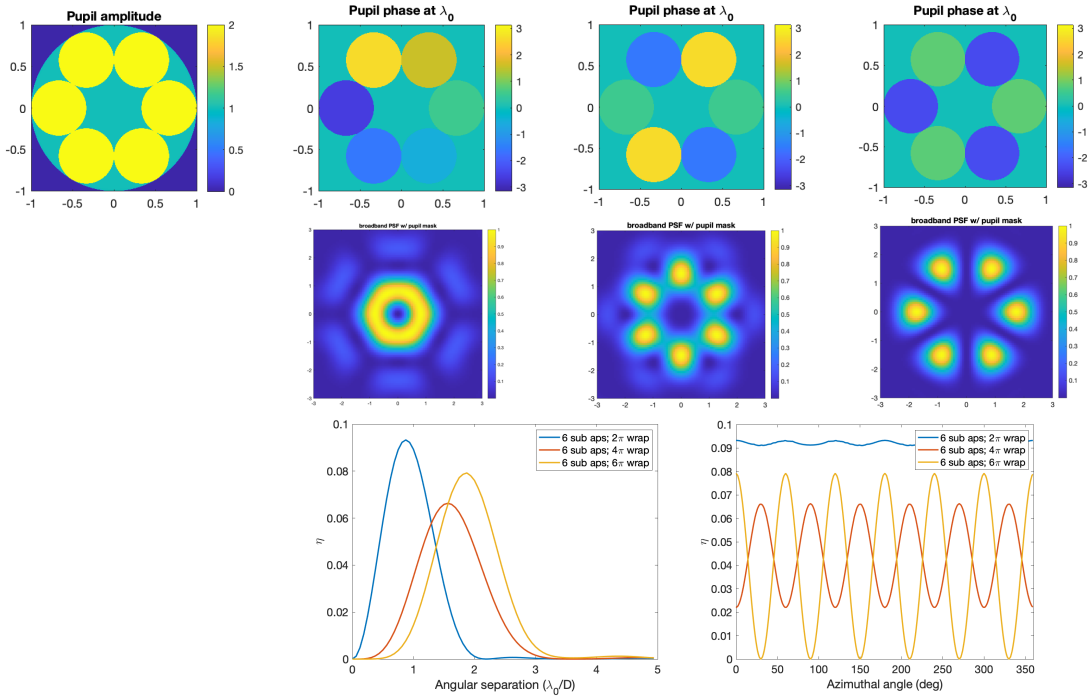

Figure 12. 6-subaperture nullers. Top row: pupil amplitudes and phases for $2\pi$, $4\pi$ and $6\pi$ phase wraps for 6 maximally-spaced subapertures. Middle row: resultant focal-plane PSFs. Bottom row: resultant radial and azimuthal efficiencies.

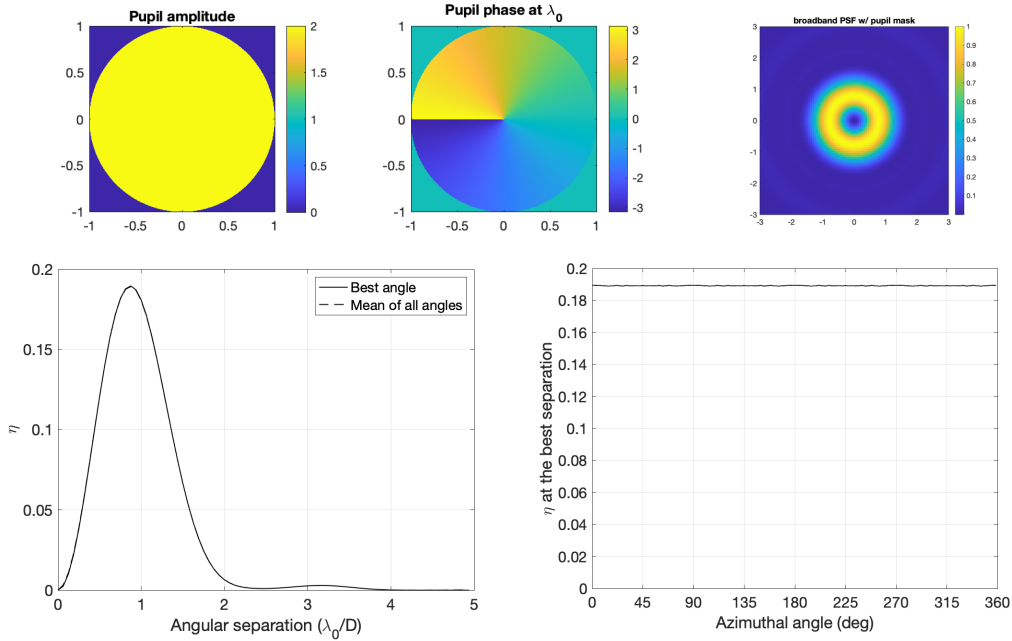

Figure 13. Vortex nuller. Top row: pupil amplitudes and phases, as well as the resultant focal-plane PSF for a charge 1 vortex. Bottom row: the resultant radial and azimuthal efficiencies.

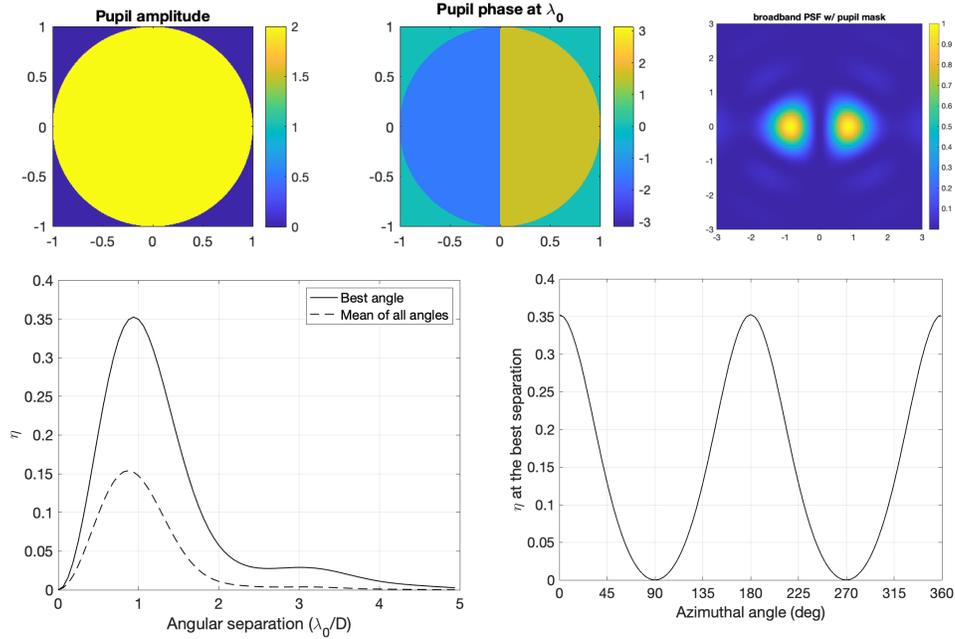

Figure 14. Phase-knife nuller. Top row: pupil amplitudes and phases, as well as the resultant focal-plane PSF for a $\pi$ phase step across the aperture centerline. Bottom row: the resultant radial and azimuthal efficiencies.

### 3.2 Summary of model results

Several results emerge from this set of model calculations. First, the IWAs of most of these cross-aperture nullers are quite small, on the order of 1/2 $\lambda/D$ or even smaller, with most efficiency curves peaking near ~ 1 $\lambda/D$. Second, several of the subaperture nullers have low peak efficiencies, on the order of 10% or less, due to their low use of telescope area. Next, as N increases, the number of possible phase wrap solutions, n, increases, as mentioned. In analogy to the vortex nuller, we can call n the "charge" of a "circular" nuller (with as mentioned, n any integer that satisfies $1 \leq n \leq N/2$). Next, the azimuthal response present in the different PSFs is quite varied, both for different N, and for different n for a given N. The vortex nuller is the most extreme case, with no azimuthal variation at all. Indeed, as N increases, the $2\pi$ phase wrap solution begins to resemble the vortex solution more and more.

Interestingly, the highest efficiency curve is found not for any of the circular nullers, but instead for the phase-knife nuller, which has a peak efficiency of 35.2%. This is followed in second place by the vortex nuller, with an azimuthally uniform peak efficiency of 19.0%. In contrast, the phase-knife nuller has a strong azimuthal response curve. On the other hand, the average azimuthal response of the phase-knife nuller is roughly the same as the constant azimuthal response of the vortex nuller. Thus, as rotating the phase knife to an orthogonal location would double the observation time with the phase-knife, spectroscopy on sources without a known azimuth is likely best carried out with a vortex nuller, while for exoplanets with known azimuths, a properly-oriented phase knife would yield roughly twice the signal to noise ratio (SNR) of the vortex nuller, thus allowing the phase knife coronagraph to acquire spectra of known targets in ¼ the time of a vortex nuller.

Finally, note that in the phase-knife case, the two subapertures are not identical, but are instead mirror images of each other, which breaks the usual design rule of interferometric arrays (identical subapertures are consistent with the array theorem). However, the use of mirror-image subapertures is allowed here because of the use of single mode fibers, as mirror-symmetric subapertures couple to a single-mode fiber with equal efficiency. Note that even though the baseline between subapertures is not well defined in the usual interferometric sense for mirror-symmetric subapertures, the baseline is not a quantity of astrophysical interest, as we care only about maximizing exoplanet transmission efficiency.

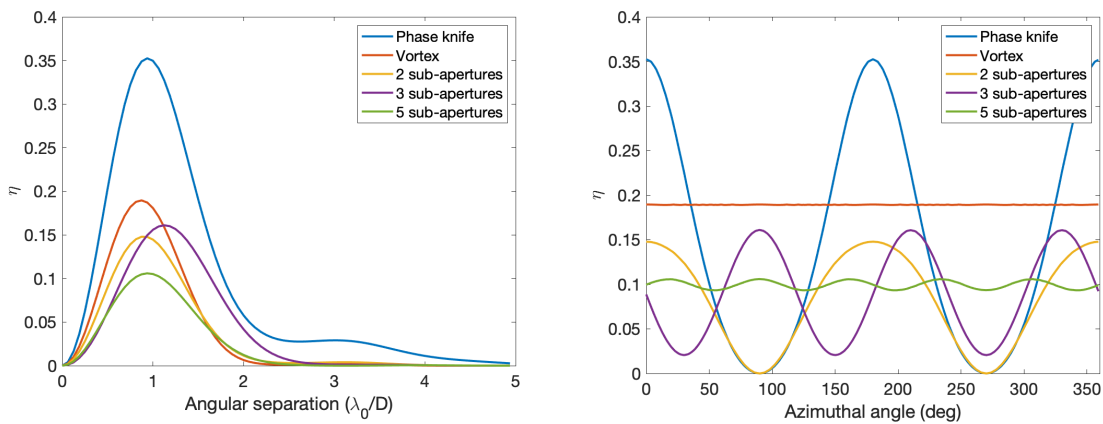

Figure 15. Radial (left) and azimuthal (right) efficiency curve comparison for the phase knife, vortex and maximally-spaced 2, 3, and 4 subaperture nullers..

## 4. SUMMARY

A systematic comparison of a variety of nulling coronagraph configurations has shown that single-mode nulling coronagraphs all have very small IWAs, i.e., well under $\lambda/D$, and that subaperture-based nullers have lower efficiencies than those that make use of the entire telescope aperture, such as the phase-knife and vortex nullers. As a result, the IWA in isolation can be a misleading quantity, as nullers with higher peak transmissions and larger IWAs can have higher absolute throughputs at all angles than a nulling coronagraph with a smaller formal IWA but low intrinsic efficiency.

The single-mode nulling configuration with the highest efficiency has been found to be the phase-knife nuller, followed by the vortex nuller, with a peak efficiency lower by a factor of almost 2. Of these, the phase-knife nuller shows a strong azimuthal response, allowing azimuthal source position determination, while the vortex nuller has no azimuthal response variation at all. Thus, if one knows where an exoplanet is, use of a phase-knife nuller provides the highest signal to noise ratio in a given time, whereas the use of a vortex nuller allows for spectroscopy even without knowing at what azimuth an exoplanet is located. Both of these nulling coronagraphs thus have useful roles to play.

**REFERENCES**

[1] Habex report: https://www.jpl.nasa.gov/habex/documents/ (2019).
[2] LUVOIR report: https://asd.gsfc.nasa.gov/luvoir/reports/ (2019).
[3] Serabyn, E. "Nulling Interferometry," in *The WSPC Handbook of Astronomical Instrumentation*, ed. D. N. Burrows, World Scientific (2021).
[4] Serabyn, E., Mennesson, B., Martin, S., Liewer, K. & Kühn, J., M.N.R.A.S. 489, 1291 (2019).
[5] Hinz, P. M., Angel, J. R. P., Hoffmann, W. F., McCarthy Jr, D. W., McGuire, P. C., Cheselka, M., Hora J. L. & Woolf, N. J., Nature 395, 251 (1998).
[6] Baudoz, P. et al., Proc. SPIE vol. 3353, https://doi.org/10.1117/12.321671 (1998).
[7] Serabyn, E., Mennesson, B., Colavita, M.M., Koresko, C. & Kuchner, M. J., ApJ 748, 55 (2012).
[8] Ertel, S. et al., Astron. J. 159, 177 (2020).
[9] Ruane, G., Wang, J., Mawet, D., Jovanovic, N., Delorme, J.-R., Mennesson, B. & Wallace, J. K., ApJ 867, 143 (2018)
[10] Echeverri, D. et al., in Proc. SPIE vol. 11446, 1144619 (2020).
[11] Echeverri, D. et al., in Proc. SPIE vol. 11823, 118230A (2021).
[12] Beichman, C. A., Woolf, N. J. & Lindensmith, C., "*The Terrestrial Planet Finder (TPF): A NASA Origins program to search for habitable planets*", JPL publ. 99-3 (1999).
[13] Mennesson, B., Leger, A. & Ollivier, M., Icarus 178, 570 (2005).
[14] Velusamy, T., Beichman, C.A. & Shao, M., in *Optical and IR Interferometry from Ground and Space*, ASP Conf. Ser. Vol 194, 430 (1999).
[15] Angel, J. R. P & Wolf, N. J., ApJ 475, 373 (1997).
[16] Karlsson, A., Wallner, O., Armengol, J.P. & Absil, O., Proc. SPIE 5491, 831 (2004).
[17] Serabyn, G., Proc. SPIE 5491, 1639 (2004).
[18] Angel, J. R. P., in The Next Generation Space Telescope, ed. P. Bely & C. J. Burrows (Baltimore : Space Telescope Science Institute), 81 (1990).
[19] S. R. Martin & A. J. Booth, A&A 520, A96 (2010).
[20] Roddier, F. & Roddier, C., P.A.S.P. 109, 815 (1997).